\documentclass[fleqn,twoside]{article}
\usepackage{espcrc2}
\usepackage{epsfig}
\usepackage{graphicx}

\def\lsi{\raise0.3ex\hbox{$<$\kern-0.75em\raise-1.1ex\hbox{$\sim$}}}
\def\gsi{\raise0.3ex\hbox{$>$\kern-0.75em\raise-1.1ex\hbox{$\sim$}}}

\title{
\vspace*{-1.2cm}
\mbox{} \hfill {\small ITP-Budapest 587, DESY 02-140}\\
\vspace*{0.7cm}
The QCD equation of state at finite T and $\mu$}
\author{
F. Csikor\address[ELTE]{Institute for Theoretical Physics, E\"otv\"os University,
P\'azm\'any P. 1/A, H-1117 Budapest, Hungary}, G.I. Egri\addressmark[ELTE], 
Z. Fodor\addressmark[ELTE],
S.D. Katz\address{Deutsches Elektronen-Synchrotron DESY, Notkestr. 85, D-22607,
Hamburg, Germany}\thanks{on leave from
Institute for Theoretical Physics, E\"otv\"os University,
P\'azm\'any P. 1/A, H-1117 Budapest, Hungary}, K.K. Szab\'o 
and A.I. T\'oth\addressmark[ELTE]}

\begin{document}

\begin{abstract}
We calculate the pressure (p), the energy density ($\epsilon$) 
and the baryon density ($n_B$) of QCD at
finite temperatures (T) and
chemical potentials ($\mu$). The recently proposed overlap improving multi-parameter
reweighting technique is used to determine observables at nonvanishing
chemical potentials. Our results
are obtained by studying $n_f$=2+1 dynamical
staggered quarks with semi-realistic masses on $N_t=4$ lattices. 
\end{abstract}

\maketitle

\begin{figure}
\begin{center}
\includegraphics*[width=6.9cm,bb=0 100 413 373]{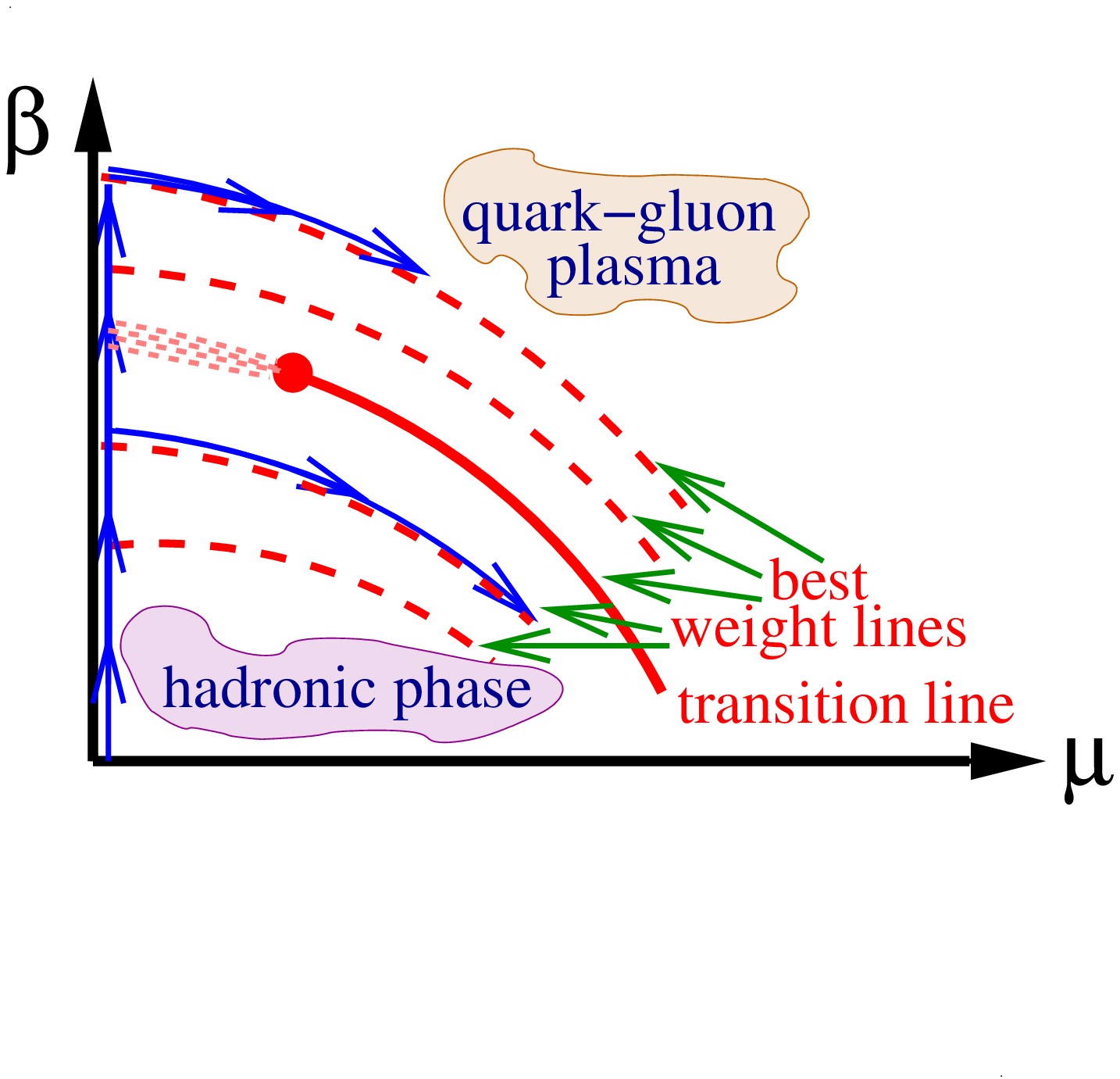}
\vspace{-1.3cm}
\end{center}
\caption{\label{weightlines}
The best weight lines on the $\mu$--$\beta$ plane. 
In the middle we 
indicate the transition line. Its first dotted part
is the crossover region. The blob represents the 
critical endpoint, after which the transition is of first order. 
The integration paths used to calculate $p$ are shown by the 
arrows along the $\beta$ axis and the best weight lines.
}
\vspace{-0.6cm}
\end{figure}

QCD at nonzero density  is
easily formulated on the lattice
by multiplying the forward/backward links by 
$\exp(\pm \mu)$.
However, standard importance sampling based 
Monte-Carlo techniques can not be used 
at $\mu\neq$0.  
Up to now, no technique was suggested capable of giving
the equation of state (EOS)
at $\mu$$\neq$0, which is essential 
to describe the quark gluon plasma (QGP) formation
at heavy ion collider experiments. Results are only available for $\mu$=0
(e.g. \cite{Gottlieb:1996ae,Karsch:2000ps,AliKhan:2001ek}) at
T$\neq$0.

\begin{figure}
\begin{center}
\includegraphics*[width=6.9cm,bb=0 280 570 700]{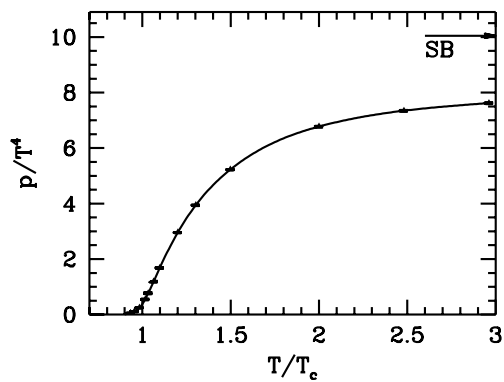}
\vspace{-1.3cm}
\end{center}
\caption{\label{eos_p0}
$p$ normalised by $T^4$ as a function of $T/T_c$ at $\mu=0$ 
The $N_t$=4 SB limit is also shown.
}
\vspace{-0.6cm}
\end{figure}

The overlap improving
multi-parameter reweighting \cite{Fodor:2001au} opened the possibility
to study lattice QCD
at nonzero $T$ and $\mu$. First one produces
an ensemble of QCD configurations at $\mu$=0 and T$\neq$0.
Then reweighting factors
are determined
at $\mu\neq 0$ and at a lowered T. The idea is 
expressed in terms of the partition function
\begin{eqnarray}\label{reweight}
Z(\beta,\mu,m) = \int {\cal D}U\exp[-S_{g}(\beta)]\det M(\mu,m) 
\nonumber \\
=\int {{\cal D}U \exp[-S_{g}(\beta_0)]\det M(\mu=0,m)}
 \\
{ \left\{\exp[-S_{g}(\beta)+S_{g}(\beta_0)]
\frac{\det M(\mu,m)}{\det M(\mu=0,m)}\right\} },
\nonumber\end{eqnarray}
where $m$ is the quark mass, $S_g$ is the action of the gluonic field ($U$), 
at $\beta=6/g^2$ coupling. 
At nonzero 
$\mu$ one gets a complex fermion determinant $\det M$ 
which 
spoils importance sampling.
Thus, the first line of eq. (\ref{reweight}), $\mu\neq$0,
is rewritten in a way that the second line of eq. (\ref{reweight})
is used as an integration measure (at $\mu$=0 with 
importance sampling) and the curly
bracket is measured on each independent configuration and is interpreted 
as a weight factor $\{ w(\beta,\mu,m,U)\}$. 
In order to maximise the accuracy of 
$Z$ the reweighting is performed along the best weight lines on the 
$\mu$--$\beta$ plane. 
These best weight lines are defined by minimising the spread of $\log w$. 

\begin{figure}
\begin{center}
\includegraphics*[width=6.9cm,bb=0 280 570 700]{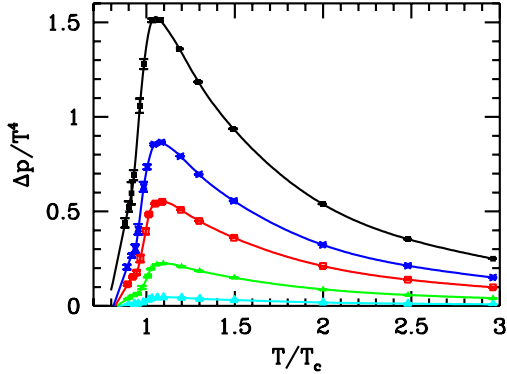}
\vspace{-1.3cm}
\end{center}
\caption{\label{eosmu_p_sub}
$\Delta p=p(\mu\neq 0,T)-p(\mu=0,T)$ normalised by $T^4$
as a function of $T/T_c$ 
for $\mu_B$=100, 210, 330, 410~MeV and
$\mu_B$=530~MeV (from bottom to top). 
}
\vspace{-0.6cm}
\end{figure}

Using the above technique, transition (or hadronic/QGP) 
configurations are reweighted to transition
(or hadronic/QGP) configurations as illustrated by Fig. \ref{weightlines}.
The technique works for
temperatures at, below and above the transition 
temperature ($T_c$). 
By using the reweighting technique, 
the phase diagram \cite{Fodor:2001au} and the location
of the critical endpoint \cite{Fodor:2001pe} was given.
Using a Taylor expansion around $\mu$=0, T$\neq$0 for small $\mu$ 
can be used
to determine thermal properties \cite{Allton:2002zi}. 
A 
different method, analytic continuation from imaginary
$\mu$, confirmed results of \cite{Fodor:2001au,Fodor:2001pe}
on the $\mu$--$T$ 
diagram \cite{deForcrand:2002ci}. 

We use 4 $\cdot N_s^3$ lattices at $T$$\neq$0 
with $N_s$=8,10,12 for reweighting 
and we extrapolate to V$\rightarrow$$\infty$ 
using the available volumes ($V$). 
At $T$=0 lattices of $24\cdot14^3$ 
are taken for vacuum subtraction and
to connect lattice parameters to physical
quantities. 14 different $\beta$ values are used, which
correspond to $T/T_c=0.8,\dots,3$. Our T=0 simulations 
provided $R_0$ and $\sigma$. The lattice spacing at $T_c$ 
is $\approx$0.25--0.30~fm. We use 2+1 flavours of
dynamical staggered quarks. 
While varying $\beta$ (thus the temperature) we keep the physical
quark masses constant at $m_{ud} \approx 65$~MeV and $m_s \approx 135$~MeV
(the pion to rho mass ratio is $m_\pi/m_\rho\approx$0.66).

\begin{figure}
\begin{center}
\includegraphics*[width=6.9cm,bb=0 280 570 700]{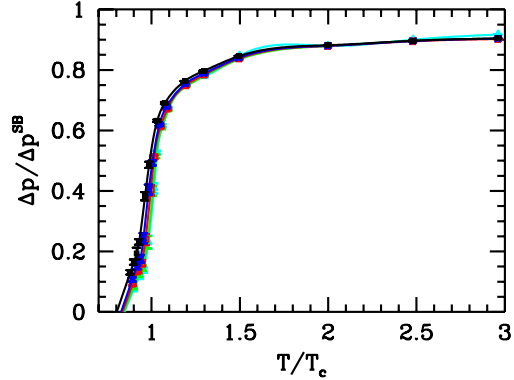}
\vspace{-1.3cm}
\end{center}
\caption{\label{eosmu_p_SB}
$\Delta p$ of the QCD
plasma normalised by $\Delta p$ of the free gas 
(SB) as a function of $T/T_c$ 
for the same $\mu_B$ values as in Fig. 3. 
}
\vspace{-0.6cm}
\end{figure}

The determination of the equation of state at $\mu\neq$0
needs several observables ${\cal O}$
at non-vanishing $\mu$ values. This can be calculated by using  
the weights of eq. (\ref{reweight})
\begin{equation}
{\overline {\cal O}}(\beta,\mu,m)=\frac{\sum \{w(\beta,\mu,m,U)\} 
{\cal O}(\beta,\mu,m,U)}{\sum
\{w(\beta,\mu,m,U)\}}.
\end{equation}

$p$ can be obtained from the partition function
as $p$=$T\cdot\partial \log Z/ \partial V$ which can be written as
$p$=$(T/V) \cdot \log Z$ for large homogeneous systems.
On the lattice we can only determine the derivatives of $\log Z$ with respect
to the parameters of the action ($\beta, m, \mu$).
Using the following notation 
$\langle  {\cal O}(\beta,\mu,m) \rangle$= 
${\overline {{\cal {O}}}(\beta,\mu,m)}_{T\neq0}-
{\overline {{\cal O}}(\beta,\mu=0,m)}_{T=0}$. 
$p$ can be written as 
an integral \cite{Engels:1990vr}:
\begin{eqnarray}
&&\frac{p}{T^4}=\frac{1}{T^3 V} \int d(\beta, m,\mu ) 
\\
&&\left(
\left\langle \frac{\partial(\log Z)}{\partial \beta}\right\rangle,
\left\langle \frac{\partial(\log Z)}{\partial m}\right\rangle,
\left\langle \frac{\partial(\log Z)}{\partial \mu }\right\rangle\right).
\nonumber
\end{eqnarray}
The integral is by definition independent of the integration path.
The chosen integration paths are shown on Fig \ref{weightlines}. 

The energy density can be written as 
$\epsilon =(T^2/V)\cdot \partial(\log Z)/\partial {T} 
+(\mu T/V)\cdot \partial(\log Z)/\partial\mu$.
By changing the lattice spacing $T$ and $V$ are simultaneously varied.
The special combination $\epsilon-3p$ contains only 
derivatives with respect to $a$ and $\mu$:
\begin{equation}
\frac{\epsilon-3p}{T^4}=-\left.\frac{a}{T^3V}\frac{\partial \log(Z)}{\partial a}\right|_\mu
+\left. \frac{\mu}{T^3 V}\frac{\partial(\log Z)}{\partial\mu}\right|_a.
\end{equation}

\begin{figure}
\begin{center}
\includegraphics*[width=6.9cm,bb=0 280 570 700]{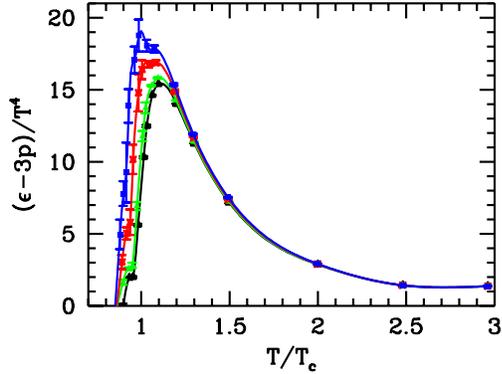}
\vspace{-1.3cm}
\end{center}
\caption{\label{interaction}
$(\epsilon-3p)/T^4$ at $\mu_B$=0, 210, 410~MeV and
530~MeV versus $T/T_c$
(from bottom to top). 
}
\vspace{-0.6cm}
\end{figure}

The quark number density is $n=(T/V)\cdot \partial \log(Z)/\partial \mu$ which 
can be measured directly or obtained from $p$ 
(baryon density is $n_B$=$n$/3 and baryonic chemical 
potential is $\mu_B$=3$\mu$).

We present direct lattice results on $p(\mu=0,T)$, 
$\Delta p(\mu,T)=p(\mu\neq 0,T)-p(\mu=0,T)$, $\epsilon(\mu,T)$-3$p(\mu,T)$ and 
$n_B(\mu,T)$. Note, that in \cite{Csikor:2002aa} additional overall factors
were used to help the phenomenological interpretation.
Our statistical errorbars are rather small, 
sometimes smaller than the thickness of the lines. 
 
\begin{figure}
\begin{center}
\includegraphics*[width=6.9cm,bb=0 280 570 700]{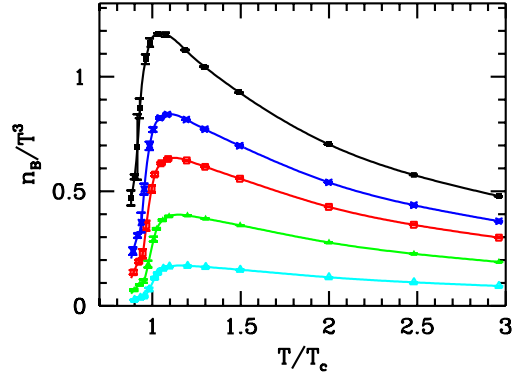}
\vspace{-1.3cm}
\end{center}
\caption{\label{density}
$n_B/T^3$ versus $T/T_c$ 
for the same $\mu_B$ values as in Fig. 3 
(from bottom to top).
}
\vspace{-0.6cm}
\end{figure}

Fig. \ref{eos_p0} shows $p$  at $\mu$=0. 
On Fig. \ref{eosmu_p_sub} we present $\Delta p/T^4$ for five different 
$\mu$ values. 
Fig. \ref{eosmu_p_SB} gives $\Delta p(\mu,T/T_c)$ 
normalised by $\Delta p^{SB}\equiv \Delta p(\mu,T\rightarrow\infty)$.
Notice the interesting scaling behaviour. 
$\Delta p / \Delta p^{SB}$ depends only on T and it
is practically independent of $\mu$ in the analysed region. 
Fig. \ref{interaction} shows $\epsilon$-3$p$ normalised by
$T^4$, which tends to zero for large $T$. 
Fig. \ref{density} gives the baryonic density
as a function of $T/T_c$ for different $\mu$-s. As it can be seen 
the densities exceed the nuclear density
by up to an order of magnitude.

An important finding concerns  
the applicability of our reweighting
method: the maximal $\mu$  
scales with the volume as 
$\mu_{\rm{max}}\cdot a \sim (N_t\cdot N_s^3)^{-0.25}$.
If this behaviour persists, one could --in principle--
approach the true continuum limit   
($a \sim 1/N_t \sim (N_t\cdot N_s^3)^{-0.25}$, thus
$\mu_{\rm{max}}$$\approx$const.).

Future analyses should
be done at smaller lattice spacings and quark masses.
A detailed version of this work can be found 
elsewhere \cite{Csikor:2002aa}.

{\bf Acknowledgements:} 
This work was partially supported by Hungarian Scientific
grants OTKA-T37615/\-T34980/\-T29803/\-M37071/\-OMFB1548/\-OMMU-708. 
For the simulations a modified version of the MILC
public code was used (see http://physics.indiana.edu/\~{ }sg/milc.html). 
The simulations were carried out on the 
E\"otv\"os Univ., Inst. Theor. Phys. 163 node parallel PC cluster.

\end{document}